# Role of structural factors in formation of chiral magnetic soliton lattice in $Cr_{1/3}NbS_2$

**L. M. Volkova and D. V. Marinin**
*Institute of Chemistry, Far Eastern Branch of the Russian Academy of Sciences, 690022 Vladivostok, Russia*

The sign and strength of magnetic interactions not only between nearest neighbors, but also for longer-range neighbors in the $Cr_{1/3}NbS_2$ intercalation compound have been calculated on the basis of structural data. It has been found that left-handed spin helices in $Cr_{1/3}NbS_2$ are formed from strength-dominant at low temperatures AFM interactions between triangular planes of $Cr^{3+}$ ions through the plane of just one of two crystallographically equivalent diagonals of side faces of embedded into each other trigonal prisms building up the crystal lattice of magnetic $Cr^{3+}$ ions. These helices are oriented along the *c* axis and packed into two-dimensional triangular lattices in planes perpendicular to these helices directions and lay one upon each other with a displacement. The competition of the above AFM helices with weaker inter-helix AFM interactions could promote the emergence of a long-period helical spin structure. One can assume that in this case the role of Dzyaloshinskii-Moriya interaction consists in final ordering and stabilization of chiral spin helices into a chiral magnetic soliton lattice. The possibility of emergence of solitons in $M_{1/3}NbX_2$ and $M_{1/3}TaX_2$ (M = Cr, V, Ti, Rh, Ni, Co, Fe, and Mn; X = S and Se) intercalate compounds has been examined. Two important factors caused by the crystal structure (predominant chiral magnetic helices and their competition with weaker inter-helix interactions not destructng the system quasi-one-dimensional character) can be used for the crystal chemistry search of solitons.

## I. INTRODUCTION

Since the discovery of magnetic solitons, the problems of the relation of their emergence to the crystal structure continue to attract a substantial attention. The results of these studies are necessary not only for creating theoretical models of this phenomenon, but also for performing the crystal chemistry search of compounds with structural characteristics promoting the formation of magnetic solitons.

The materials with structural chirality, i.e., left or right handedness, are known to be capable to serve as a basis for the emergence of a chiral magnetic soliton lattice.[1-7] When this kind of handedness appears in the structure of atoms in a solid, it affects the way that the magnetic moments of unpaired electrons organize themselves through the relativistic spin-orbit Dzyaloshinskii-Moriya (DM) interactions.[8-11] For example, the authors of Ref. 12 managed to demonstrate, by the methods of X-ray diffraction using synchrotron radiation and polarized neutron small-angle diffraction, that left- and right-handed crystals can be found for MnSi and its iron substituted analogs. The structural chirality rigorously determines the magnetic chirality of these compounds: left-(right-)handed crystalline chirality establishes left (right) handedness of the magnetic helix.

However, in most cases, it remains unclear what specific structural fragments can participate in formation of spin helices. As was rightly mentioned in Ref. 13, in most of relevant works, chirality of the crystal structure is related to just one parameter – the absence of inversion center. Indeed, all the chiral structures are noncentrosymmetric. However, symmetry groups are chiral only when they do not contain symmetry elements of the type II (inversion center, symmetry planes, and rotoinversion axis) and belong to families of point groups of the rotating cone, twisted cylinder, or sphere with rotating surface points[14]. If these conditions are not present, the noncentrosymmetric screw spin structure has two domains: right-handed and left-handed.

It is generally accepted that in noncentrosymmetric crystals the selection of one of the domains produces the relativistic spin-orbit DM interactions and forms the chiral helimagnetic structure.[8-11] Without the DM interaction, the left-handed and right-handed chiral magnetic structures are degenerate and the magnetic structure inevitably becomes achiral. At the same time, magnetic ions even in noncentrosymmetric structure can be localized in a special position and form a centrosymmetric sublattice. For example, the crystal lattice of $Cr_{1/3}NbS_2$[15] (space group $P6_322$) is noncentrosymmetric. However, $Cr^{3+}$ ions in this structure are localized in a special position 2c (1/3, 2/3, 1/4 and 2/3, 1/3, 3/4) and form a centrosymmetric crystal sublattice, which can be presented in the space group $P\bar{1}$ with $Cr^{3+}$ ions localized in a common position 2i (x, y, z and -x, -y, -z) at preservation of the same unit cell parameters and $Cr^{3+}$ ions coordinates. Here, the centrosymmetric character of the magnetic subsystem can be disrupted by noncentrosymmetric positions of other ions in the structure. Due to their noncentrosymmetric positions, these intermediate ions ($S^{2-}$ ions in $Cr_{1/3}NbS_2$) could unequally enter local spaces of magnetic ions located in crystallographically equivalent positions and, therefore, create, at the expense of exchange interactions, the difference in magnetic moments as in the value as in the direction in the magnetic subsystem. The loss of inversion center in the magnetic subsystem of structures must release the forces of relativistic nature (DM).

According to Refs. 8-11, the DM interactions favor a screwlike arrangement of the magnetic moments, but they must compete with ferromagnetic exchange, which tries to align all the magnetic moments in the same direction. The result is a helical magnetic arrangement with a winding period of several tens of nanometers, which is much longer than the lattice constant. However, the reasons of emergence of helical structures can consist not only in relativistic spin–lattice and spin–spin interactions, but also in sharp anisotropy of exchange interaction.[16] Competing exchange magnetic interactions often lead to non-collinear magnetic structure that can be helical.[17, 18]

From the crystal chemistry point of view, one can calculate the sign and relative strength of exchange magnetic interaction on the basis of structural data and distinguish left(right) handedness of the magnetic helix in the structure. One can determine the geometric competition of magnetic interactions, which could facilitate the formation of superstructures with high period or, in opposite, disrupt this superstructure through competition in strength with helical interactions. However, it appears impossible to figure out ordering and stabilization of chiral spin helices in the form of an orderly mobile system without participation of relativistic forces. The soliton dynamics is indirectly related to the crystal structure. As was demonstrated in Ref. 1, only in quasi-one-dimensional systems a soliton behaves as a mobile quasi-particle and can manifest itself in neutron quasi-elastic scattering and thermodynamics of such systems.

The structure of crystal sublattice of magnetic ions serves as a basis, whose filling with various additional details such as intermediate ions and conductivity electrons and vacancies as well as changing the ion sizes can provide different magnetic properties of the substance. Although the relation between the spin chirality and the crystal structure has been established experimentally, it is difficult to single out specific features of this relation for an individual compound.

In the present work, we will try to reveal in detail the role of structural factors in the emergence of solitons in the layered intercalation compound $Cr_{1/3}NbS_2$. This compound was investigated in many works. Magnetization measurements and small angle neutron scattering experiments demonstrate that below 127 K $Cr_{1/3}NbS_2$ is a long-period left-handed helimagnet, whose magnetic moments spiral along the hexagonal *c*-axis with a pitch of 48 nanometers.[19-22] As was shown in Ref. 20, the reason of emergence of a long period helical spin structure in $Cr_{1/3}NbS_2$ at low temperatures consists in antisymmetric exchange interaction, whereas it was not established what magnetic ions participated in this interaction. The authors of Ref. 23 determined the temperature and magnetic field dependence of the magnetic, transport, and thermal properties of $Cr_{1/3}NbS_2$ single crystals. Using the Lorenz microscopy and small-angle electron diffraction, the authors of Ref. 4 directly presented how the chiral magnetic soliton lattice continuously evolved from a chiral helimagnetic structure in small magnetic fields in $Cr_{1/3}NbS_2$. Moreover, the dynamics of soliton transformation into a conventional ferromagnetic phase under magnetic field effect directed perpendicularly to the chiral axis was established and, thus, the possibility to control twisting of spin helices by an external magnetic field was demonstrated.

The objective of the present work was to search, on the basis of structural data, specific magnetic couplings (MC) forming left-handed magnetic helices and their competition with other MC that would facilitate the emergence of the chiral helimagnetic structure in the $Cr_{1/3}NbS_2$ intercalate  The implementation of this objective will serve as a proof that above the transition the magnetic system already contains a nucleus of the system (determined by the crystal structure of $Cr_{1/3}NbS_2$), into which it can transform with participation of DM relativistic forces. Besides, we will examine the possibility of emergence of solitons in 16 more representatives of this type of intercalation compounds: $M_{1/3}NbS_2$ (M = V, Ti, Ni, Co, Fe, and Mn), $M_{1/3}NbSe_2$ (M = Cr, Rh, V, Ti, and Co), $M_{1/3}TaX_2$ (M = Cr, V and, Rh), and $M_{1/3}TaSe_2$ (M = Cr and V) on the basis of comparison of magnetic coupling parameters in these compounds with respective ones in $Cr_{1/3}NbS_2$. For this purpose, we will calculate, on the basis of structural data, the sign and strength of magnetic couplings in the compounds under examination not only between nearest neighbors, but also at long distances (because in this case they have a crucial role) and analyze their competition.

## II. METHOD OF CALCULATION

To determine the characteristics of magnetic interactions (type of the magnetic moments ordering and strength of magnetic coupling) in the intercalation complexes $M_{1/3}NbX_2$ and $M_{1/3}TaX_2$ (M = d-elements, X =

S and Se), we used the earlier developed phenomenological method, which we named the *"crystal chemistry method"*, and the program "MagInter" developed on its basis.[24, 25] The method enables one to determine the sign (type) and strength of magnetic couplings on the basis of structural data. According to this method, a coupling between magnetic ions $M_i$ and $M_j$ emerges in the moment of crossing the boundary between them by an intermediate ion $A_n$ with the overlapping value of ~0.1 Å. The area of the limited space (local space) between the ions $M_i$ and $M_j$ along the bond line is defined as a cylinder, whose radius is equal to these ions radii. The strength of magnetic couplings and the type of magnetic moments ordering in insulators is determined mainly by the geometrical position and the size of intermediate $A_n$ ions in the local space between two magnetic ions $M_i$ and $M_j$. The positions of intermediate ions $A_n$ in the local space are determined by the distance $h(A_n)$ from the center of the ion $A_n$ up to the bond line $M_i$-$M_j$ and the degree of the ion displacement to one of the magnetic ions expressed as a ratio ($l_n'/l_n$) of the lengths $l_n$ and $l_n'$ ($l_n \le l_n'$; $l_n' = d(M_i - M_j) - l_n$) produced by the bond line $M_i$-$M_j$ division by a perpendicular made from the ion center.

The intermediate ions $A_n$ will tend to orient magnetic moments of $M_i$ and $M_j$ ions and make their contributions $j_n$ into the emergence of antiferromagnetic (AFM) or ferromagnetic (FM) components of the magnetic interaction in dependence on the degree of overlapping of the local space between magnetic ions ($\Delta h(A_n)$), asymmetry ($l_n'/l_n$) of position relatively to the middle of the $M_i$-$M_j$ bond line, and the distance between magnetic ions ($M_i$-$M_j$).

Among the above parameters, only the degree of space overlapping between the magnetic ions $M_i$ and $M_j$ ($\Delta h(A_n) = h(A_n) - r_{A_n}$) equal to the difference between the distance $h(A_n)$ from the center of $A_n$ ion up to the bond line $M_i$-$M_j$, and the radius ($r_{A_n}$) of the ion $A_n$ determined the sign of magnetic interaction. If $\Delta h(A_n)$<0, the $A_n$ ion overlaps (by |$\Delta h$|) the bond line $M_i$-$M_j$ and initiates the emerging contribution into the AFM-component of magnetic interaction. If $\Delta h(A_n)$>0, there remains a gap (the gap width $\Delta h$) between the bond line and the $A_n$ ion, and this ion initiates a contribution to the FM-component of magnetic interaction.

The sign and strength of the magnetic coupling $J_{ij}$ is determined by the sum of the above contributions:

$$J_{ij} = \sum_n j_n .$$

The value $J_{ij}$ is expressed in units of Å$^{-1}$. If $J_{ij} < 0$, the type of $M_i$ and $M_j$ ions magnetic ordering is AFM and, in opposite, if $J_{ij} > 0$, the ordering type is FM.

The value of the contribution to the AFM or FM coupling components is maximal, if the intermediate ion position is in the central one-third of the local space between magnetic ions. For the maximal contribution into the AFM-component of the coupling, the intermediate ion must be at the closest distance from the axis, while in the case of the FM-component, in opposite, from the surface of the cylinder limiting the space area between the magnetic ions. In the absence of direct magnetic interaction, the distance between the magnetic ions $M_i$ and $M_j$ affects only the value of contribution, but does not determine its sign. We assumed in our calculations the coupling strength to be inverse-proportional to the square of the distance between the magnetic ions $M_i$ and $M_j$. However, the dependence of the coupling strength on the above distance is more complicated. Along with the distance increase, the coupling strength decrease occurs at a higher rate. Juxtaposition of the data obtained using our method with experimental results of studies of the known magnetic compounds showed that the coupling strength was inverse-proportional to the distance square at the d($M_i$-$M_j$) distance increase up to ~8 Å, while during further increase of the distance the coupling strength must be inverse-proportional not to the square, but to the cube of the distance. However, the available literature does not contain sufficient reliable data to take the above effect into account in our method. As a result, the strength of couplings between ions located at long distances might be artificially overrated.

The method is sensitive to insignificant changes in the local space of magnetic ions and enables one to find intermediate ions localized in critical positions, deviations from which would result in the change of the magnetic coupling strength or spin reorientation (AFM-FM transition, for instance, under effect of temperature or external magnetic field). In the compounds under examination, one observes three critical positions: 'a', 'c', and 'd'.[24, 25] In the 'a' position ($h(A_n) = r_M + r_{An} - 0.1$), the ion $A_n$ enters the local space by less than $\Delta a$ ~ 0.1 Å and does not initiate the emergence of magnetic interaction ($j_{A_n}$ = 0). However, at slight decrease of the distance $h(A_n)$ from the $A_n$ ion center to the bond line $M_i$-$M_j$ (the $A_n$ ion displacement inside this area), there emerges a substantial contribution of this ion to the coupling FM component. In the 'c' position ($l'_n/l_n = 2$), the $A_n$ ion is located at the boundaries of the central one-third of the space between magnetic ions. In this case, the insignificant displacement of the $A_n$ ion to the center in parallel to the bond line $M_i$-$M_j$ line results in a dramatic increase of the magnetic interaction strength. The position 'd' emerges in the case when several intermediate ions $A_n$ are located between the magnetic ions $M_i$ and $M_j$ and the sums of these ions contributions $j_n$ to AFM and FM components are approximately equal making the magnetic coupling weak and unstable. Small displacement of any of the intermediate ions might result in the coupling complete disappearance or AFM-FM transition.

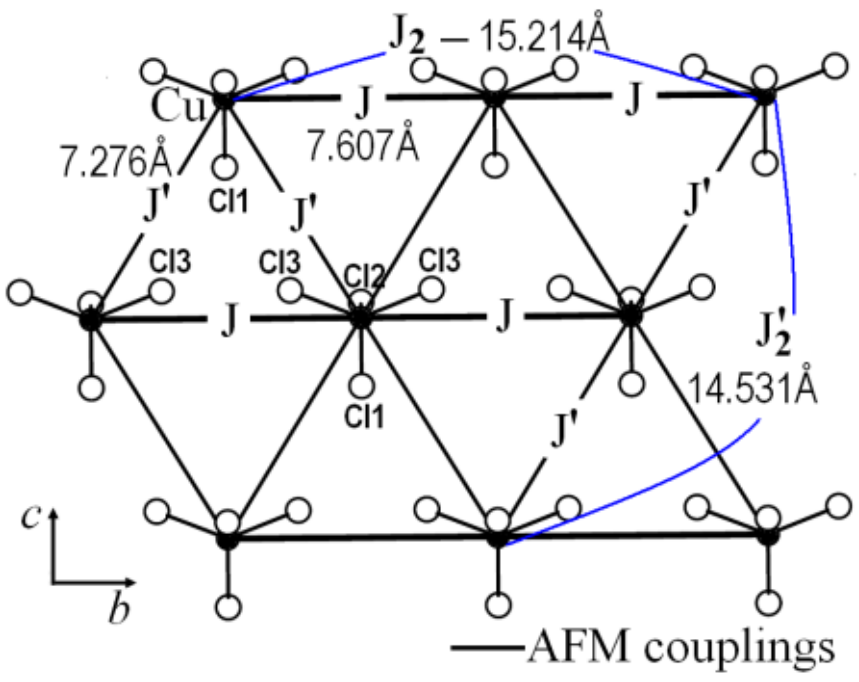

FIG. 1.Triangular magnetic planes with exchange couplings of *J*, *J'*, $J_2$ and $J'_2$ in antiferromagnet $Cs_2CuCl_4$[26]

Let us illustrate the viability of the crystal chemical method on the example of evaluation of exchange parameters in isostructural antiferromagnets $Cs_2CuCl_4$[26] and $Cs_2CuBr_4$[27] with anisotropic triangular lattice of spin-1/2 $Cu^{2+}$ ions. According to our calculations, in $Cs_2CuCl_4$ the ratio $J'/J$ = 0.39 of the two AFM exchange constants *J* (*J* = -0.0100 $Å^{-1}$, d(Cu-Cu) = 7.607 Å) and *J'* (*J'*= -0.0039 $Å^{-1}$, d(Cu-Cu) = 7.276 Å) in the triangular lattice (Fig. 1) just slightly exceeds inelastic neutron-scattering ($J'/J$ = 0.34)[28] and electron spin resonance spectroscopy in magnetic fields results ($J'/J$ = 0.30)[29]. For the isostructural $Cs_2CuBr_4$, our result $J'/J$ = 0.46 (*J* = -0.0102 $Å^{-1}$, d(Cu-Cu) = 7.965 Å) and *J'* (*J'*= -0.0047 $Å^{-1}$, d(Cu-Cu) = 7.606 Å) is also in agreement with the experimental value $J'/J$ = 0.41[29] determined by the electron spin resonance spectroscopy, but differs significantly from the value $J'/J$ = 0.74 found through inelastic neutron scattering experiments in Ref. 30. It is worth mentioning that, according to our calculations, in $Cs_2CuBr_4$ such value ($J'_2/J_2$ = 0.74) is obtained for next-nearest-neighbor $J_2$ ($J_2$ = -0.0151 $Å^{-1}$, d(Cu-Cu) = 15.930 Å) and $J'_2$ ($J'_2$ = -0.0112 $Å^{-1}$, d(Cu-Cu) = 15.191 Å) couplings in linear chains along respective triangle sides (Fig. 1), not for the ratio between nearest neighbors in the triangular lattice. About the same ratio $J'_2/J_2$ = 0.70 ($J_2$ = -0.0148 $Å^{-1}$, d(Cu-Cu) = 15.214 Å and $J'_2$ = -0.0104 $Å^{-1}$, d(Cu-Cu) = 14.531 Å) was obtained for $Cs_2CuCl_4$.

As initial data for calculations of the sign and strength of magnetic interactions, we used crystallographic parameters and coordinates of intercalate atoms: $M_{1/3}NbS_2$ (M = Cr[15], V[31], Ti[31], Ni[32], Co[33], Fe[34] and Mn[32]), $M_{1/3}NbSe_2$ (M = Cr[35], Rh [[15]], V[35], Ti[15] and Co[35]), $M_{1/3}TaS_2$ (M = Cr[15], V[15] and Rh[15]), and $M_{1/3}TaSe_2$ (M = Cr[15] and V[15]) at room temperature and ion radii determined in Ref. 36. Unfortunately, the crystal structures of all these compounds (aside from $Fe_{1/3}NbS_2$[34]) are not determined with proper accuracy. One requires extra correction of the general position of S(Se) and the occupancy of all atoms positions, since it greatly affects magnetic properties and, thus, interpretation of relation between them and structure and the results of calculations of the sign and strength of magnetic couplings ($J_n$) using the crystal chemistry method. In 17 intercalates, the parameters of all magnetic interactions between $Cr^{3+}$ ions were calculated (*J*1, *J*2, *J*3, *J*4, *J*5, $J_c$, and *J*6(*J*6')): starting from *J*1 couplings at shortest distances (d(Cr-Cr) = 5.7 - 6.0) along triangular sides to interactions in spirals at long distances (d(Cr-Cr) = 13.2 - 14.2). Table 1 (section III) shows room-temperature crystallographic characteristics, parameters of magnetic couplings ($J_n$) and respective distances between $Cr^{3+}$ ions as well as structural data calculated using the initial and shifted variants of S(Se) atom coordinates only for 9 of 17 examined intercalation complexes $M_{1/3}NbX_2$ and $M_{1/3}TaX_2$ (M = d-elements, X = S and Se). Figures 2 and 4 (section III) were added with distances between $Cr^{3+}$ ions corresponding to the shown *J*n interactions.

### III. STRUCTURAL BASIS OF LEFT-HANED CHIRAL MAGNETIC HELICES AND COMPETING MAGNETIC INTERACTIONS IN $Cr_{1/3}NbS_2$

The layered compound $Cr_{1/3}NbS_2$[15] (Fig. 2(a)) crystallizes in a noncentrosymmetric hexagonal space group *P*$6_3$22 No. 182 (*a* = 5.741, *c* = 12.097 Å, *z* = 6) and is an intercalation compound. Cr atoms are located ordered in octahedral voids between sandwich layers S-Nb-S in disulfide $NbS_2$. $CrS_6$ octahedra (d(Cr-S)= 2.393 Å) are not linked to each other. According to Ref. 37, intercalate ions Cr are in the trivalent state and have localized electrons with spins of *S* = 3/2. The magnetic properties of $Cr_{1/3}NbS_2$ are associated with localized moments on the intercalate ions $Cr^{3+}$.

The crystal lattice of magnetic $Cr^{3+}$ ions comprises flat triangular planes parallel to the *ab* plane (Fig. 2(b)). The length of sides of the Cr-Cr triangle is equal to 5.741Å. The triangular planes are located one above another at a distance of *c*/2 with displacements by *a*/3 and *b*/3 (Fig. 2 (c)). The nearest Cr-Cr distances between triangular planes are equal to 6.847 Å. The crystal lattice of magnetic ions $Cr^{3+}$ can be also represented as two identical sublattices embedded into each other: the first one is formed by ions in trigonal prisms vertices, the other one – by ions in centers (Fig. 2(d)). In the $Cr^{3+}$ lattice, one can single out elementary fragments in the form of triangular centered $Cr_7$ (Figs. 2(d) and 2(e)). According to our calculations (Table I), these prisms contain 2 strongest magnetic couplings: AFM nearest neighbor $J1_1$ ($J1_1$ = -0.054 $Å^{-1}$, d(Cr-Cr) = 5.741 Å) along the triangles' sides in triangular planes (Figs. 2(b) and 2(e)) and AFM $J6$ ($J6$ = -0.041 Å$^{-1}$, d(Cr-Cr) = 13.390 Å) along just one of two diagonals of the $Cr_7$ prism side faces (Fig. 2(e)), i.e., between triangular planes through the plane.

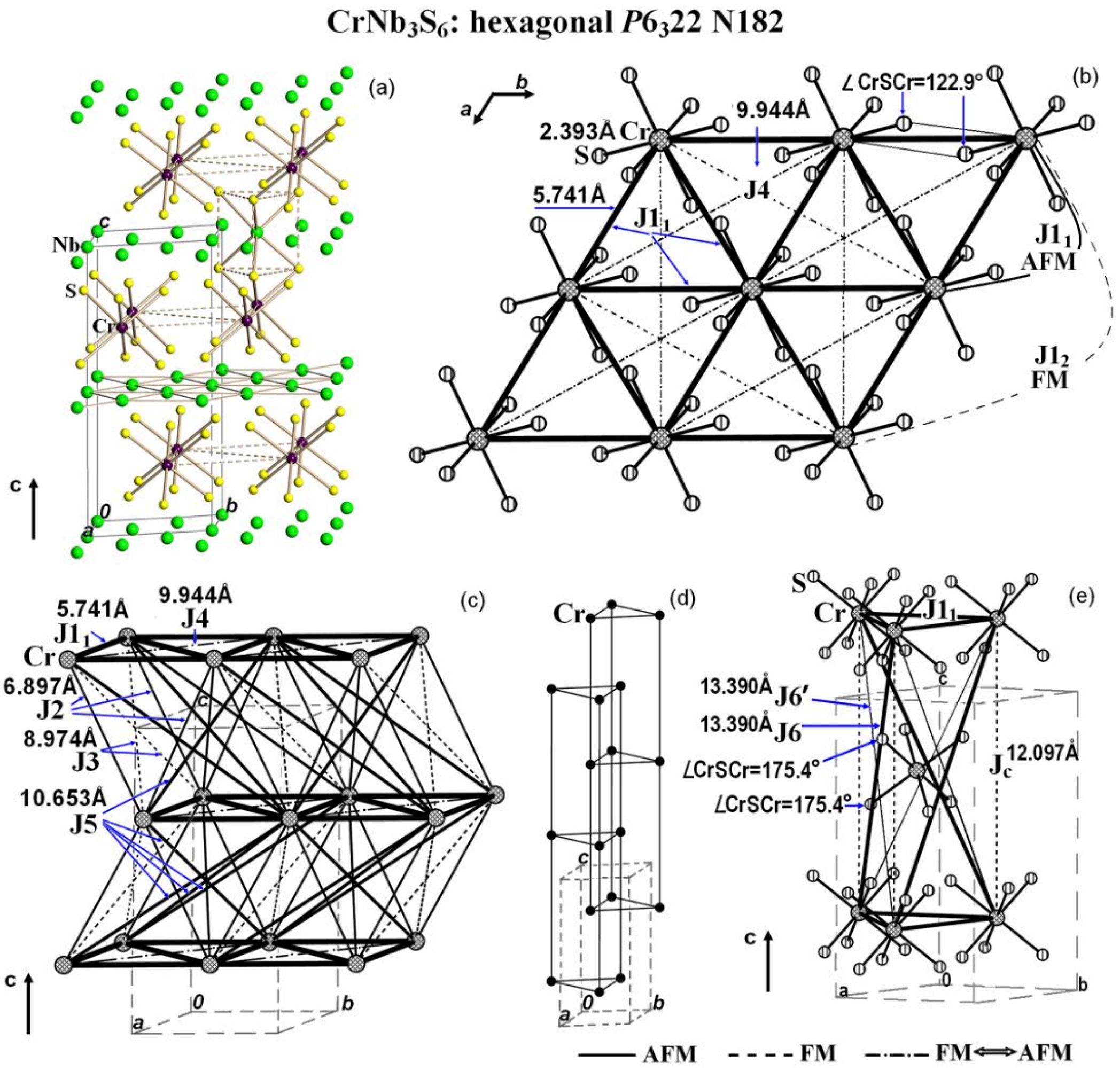

FIG. 2. The crystal structure of $Cr_{1/3}NbS_2$ (a). Triangular magnetic planes with $J1_1$, $J1_2$ and $J4$ couplings (b) and $J2$, $J3$ and $J5$ couplings between adjacent (c). Representation of the crystal structure of magnetic $Cr^{3+}$ ions as two identical sublattices embedded into each other (d). Elementary fragments as trigonal centered prisms of $Cr_7$ in the lattice of $Cr^{3+}$ and $J1_1$, $J6$, $J6$' and $J_c$ couplings (e).
In this and other figures the thickness of lines shows the strength of the $Jn$ coupling. AFM and FM couplings are indicated by solid and dashed lines, respectively. The possible FM-AFM transitions are shown by strokes in dashed lines.

The dominating AFM nearest neighbor $J1_1$ couplings initiate two S ions (angle Cr-S-Cr = 122.9°) from different $CrS_6$ octahedra, since $CrS_6$ octahedra are not linked to each other (Figs. 2(b), 3(a) and 3(b)). The contribution of each of these ions to the AFM coupling component is equal to -0.031 Å$^{-1}$. Aside from these two S ions, the local space of the $J1_1$ coupling contains four more S ions making insignificant contributions (0.002 Å$^{-1}$ each) to the coupling FM component, which just slightly decrease the latter value during the contributions summing up. Strong AFM interactions in the triangular planes are frustrated. The next-nearest-neighbor $J1_2$ ($J1_2$/$J1_1$= -0.17, d(Cr-Cr)= 11.482 Å) couplings in linear chains along the triangles sides are weak FM couplings and do not compete with AFM $J1_1$ couplings.

In the triangular planes, one observes more AFM $J4$ ($J4$/$J1_1$= 0.11, d(Cr-Cr)= 9.944 Å) couplings (Fig. 3(d)), which cold could compete with AFM $J1_1$, if they were not so weak. One should take into account that the local space of $J4$ coupling contains by less than 0.1 Å ($\Delta a$ = 0.064 and 0.067 Å) four more sulfur ions, which are simultaneously close to two critical positions ‘a’ ($\Delta a$ ~ 0.1 Å; $\Delta a = (r_M + r_{A_n}) - h_{A_n}$) and 'c' ($l'_n/l_n = 2$) (Fig. 3(a)). Below (section IV) we will demonstrate that in intercalation compounds $M_{1/3}NbX_2$ and $M_{1/3}TaX_2$ with large radius M this lead the $J4$ coupling to the AFM → FM transition and dramatic increase of the interaction

TABLE I. Selected room-temperature crystallographic characteristics, parameters of magnetic couplings ($J_n$), and structural data calculated using the initial and displaced variants of S(Se) atoms coordinates in the intercalation complexes $M_{1/3}NbX_2$ and $M_{1/3}TaX_2$ (M = d-elements, X = S and Se) (hexagonal space group $P6_322$).

| | $M_{1/3}Nb(Ta)X_2$ | $Cr_{1/3}NbS_2$ | $Cr_{1/3}TaS_2$ | $V_{1/3}TaS_2$ | $Ti_{1/3}NbS_2$ | $Cr_{1/3}NbSe_2$ | $Rh_{1/3}NbSe_2$ | $Ni_{1/3}NbS_2$ | $Co_{1/3}NbS_2$ | $Fe_{1/3}NbS_2$ |
|---|---|---|---|---|---|---|---|---|---|---|
| | References | 15 | 15 | 15 | 31 | 35 | 15 | 32 | 33 | 34 |
| | Data for ICSD | 626392[a] | 626633[a] | 651111[a] | 645331[a] | 626398[a] | 645284[a] | 42689 | 53016 | 169953 |
| | *a* (Å) | 5.741 | 5.720 | 5.727 | 5.743 | 5.976 | 5.977 | 5.758 | 5.749 | 5.7644 |
| | *c* (Å) | 12.097 | 12.128 | 12.201 | 12.416 | 12.567 | 12.459 | 11.897 | 11.886 | 12.176 |
| | Method[b], | XDS | XDS | XDS | XDS | XDP | XDS | XDP | NDP | XDS |
| | R-value[c] | | | | | | | 0.103 | 0.07 | 0.0468 |
| | M | $Cr^{+3}$ | $Cr^{+3}$ | $V^{+3}$ | $Ti^{+3}$ | $Cr^{+3}$ | $Rh^{+3}$ | $Ni^{2+}$ | $Co^{2+}$ | $Fe^{2+}$ |
| | *r* (Å) | 0.615 | 0.615 | 0.64 | 0.67 | 0.615 | 0.665 | 0.69 | 0.745 | 0.78 |
| 1NN | d(M-M) (Å) | 5.741 | 5.720 | 5.727 | 5.743 | 5.976 | 5.977 | 5.758 | 5.749 | 5.764 |
| | $J1_1^{init\ (d)}$ (Å$^{-1}$) | -0.0538 | -0.0537 | -0.0521 | -0.0475 | -0.0605 | -0.0625 | -0.0558 | -0.0549 | -0.0451 |
| | $j^{init\ (e)}$ (Å$^{-1}$) | -0.0312 | -0.0312 | -0.0304 | -0.0283 | -0.0338 | -0.0346 | -0.0321 | -0.0317 | -0.0274 |
| | angle MXM | 122.89° | 122.60° | 122.38° | 121.69° | 122.62° | 123.02° | 123.51° | 123.23° | 121.7° |
| | $J1_1^{displ\ (d)}$ (Å$^{-1}$) | -0.0041 | -0.0041 | -0.0037 | -0.0024 | -0.0068 | -0.0074 | -0.0047 | -0.0044 | -0.0018 |
| | $j^{displ\ (e)}$ (Å$^{-1}$) | -0.0062 | -0.0062 | -0.0061 | -0.0057 | -0.0068 | -0.0069 | -0.0064 | -0.0064 | -0.0055 |
| | angle MXM | 122.78°[f] | 122.49° | 122.27° | 121.58° | 122.48° | 122.89° | 123.40 | 123.13° | 121.53° |
| | $l'/l^{init\ (f)}$ | 1.990 | 1.990 | 1.989 | 1.9896 | 1.985 | 1.9850 | 1.988 | 1.988 | 1.972 |
| | $l'/l^{displ\ (f)}$ | 2.007 | 2.008 | 2.008 | 2.008 | 2.008 | 2.0070 | 2.0037 | 2.0036 | 2.0026 |
| | $\Delta l_y^{displ\ (g)}$ (Å) | 0.0060 | 0.0063 | 0.0070 | 0.0067 | 0.0100 | 0.0102 | 0.008 | 0.009 | 0.0190 |
| $1_2$NN | d(M-M) (Å) | 11.482 | 11.440 | 11.454 | 11.486 | 11.952 | 11.955 | 11.516 | 11.498 | 11.528 |
| | $J1_2^{init}$ (Å$^{-1}$) | 0.0098 | 0.0098 | 0.0100 | 0.0111 | 0.0058 | 0.0044 | 0.0082 | 0.0076 | 0.0106 |
| | $J1_2^{displ}$ (Å$^{-1}$) | -0.0032 | -0.0032 | -0.0028 | -0.0008 | -0.0084 | -0.0101 | -0.0052 | -0.0056 | -0.0013 |
| 2NN | d(M-M) (Å) | 6.897 | 6.905 | 6.939 | 7.038 | 7.169 | 7.122 | 6.814 | 6.807 | 6.938 |
| | $J2^{init}$ (Å$^{-1}$) | -0.0090 | -0.0091 | -0.0089 | -0.0083 | -0.0104 | -0.0104 | -0.0092 | -0.0093 | -0.0088 |
| | $J2^{displ}$ (Å$^{-1}$) | -0.0077 | -0.0078 | -0.0092 | -0.0086 | -0.0105 | -0.0107 | -0.0095 | -0.0095 | -0.0091 |
| 3NN | d(M-M) (Å) | 8.974 | 8.966 | 8.997 | 9.084 | 9.333 | 9.298 | 8.922 | 8.910 | 9.021 |
| | $J3^{init}$ (Å$^{-1}$) | 0.0164 | 0.0161 | 0.0163 | 0.0171 | 0.0110 | 0.0110 | 0.0162 | 0.0467 | 0.0465 |
| | $\Delta a^h$ (Å) | -0.030 | -0.028 | -0.011 | -0.005 | 0.024 | 0.085 | 0.069 | 0.134 | 0.152 |
| | $J3^{displ}$ (Å$^{-1}$) | 0.0168 | 0.0169 | 0.0167 | 0.0179 | 0.0116 | 0.0115 | 0.0165 | 0.0472 | 0.0476 |
| | $\Delta a^h$ (Å) | -0.034 | -0.032 | -0.015 | -0.010 | 0.018 | 0.079 | 0.064 | 0.129 | 0.144 |
| 4NN | d(M-M) (Å) | 9.944 | 9.907 | 9.919 | 9.947 | 10.350 | 10.352 | 9.973 | 9.958 | 9.984 |
| | $J4^{init}$ (Å$^{-1}$) | -0.0058 | -0.0058 | -0.0056 | 0.0070 | 0.0010 | 0.0004 | 0.0043 | 0.0045 | 0.0072 |
| | $\Delta a1$ (Å) | 0.064 | 0.075 | 0.091 | 0.101 | 0.103 | 0.161 | 0.147 | 0.200 | 0.203 |
| | $\Delta a2$ (Å) | 0.067 | 0.072 | 0.093 | 0.103 | 0.113 | 0.170 | 0.153 | 0.208 | 0.205 |
| | $J4^{displ}$, (Å$^{-1}$) | -0.0056 | -0.0056 | -0.0052 | -0.0068 | 0.0160 | 0.0385 | 0.0484 | 0.0266 | 0.0545 |
| | $\Delta a1$ (Å) | 0.069 | 0.068 | 0.086; | 0.096 | 0.097 | 0.155 | 0.143 | 0.195 | 0.197 |
| | $\Delta a2$ (Å) | 0.072 | 0.070 | 0.088 | 0.099 | 0.107 | 0.164 | 0.148 | 0.203 | 0.199 |
| 5NN | d(M-M), (Å) | 10.653 | 10.635 | 10.665 | 10.747 | 11.082 | 11.053 | 10.618 | 10.604 | 10.705 |
| | $J5^{init}$,(Å$^{-1}$) | -0.0174 | -0.0175 | -0.0172 | -0.0165 | -0.0197 | -0.0199 | -0.0177 | -0.0185 | -0.0180 |
| | $J5^{displ}$ (Å$^{-1}$) | -0.0170 | -0.0171 | -0.0171 | -0.0163 | -0.0195 | -0.0194 | -0.0180 | -0.0182 | -0.0178 |
| *c*NN | d(M-M)=*c*(Å) | 12.097 | 12.128 | 12.201 | 12.416 | 12.567 | 12.459 | 11.897 | 11.886 | 12.176 |
| | $J_c^{init}$ (Å$^{-1}$) | 0.0070 | 0.0063 | 0.0065 | 0.0067 | 0.0012 | 0.0013 | 0.0080 | 0.0078 | 0.0073 |
| | $J_c^{displ}$ (Å$^{-1}$) | 0.0075 | 0.0068 | 0.0069 | 0.0072 | 0.0018 | 0.0018 | 0.0084 | 0.0082 | 0.0081 |
| 6NN | d(M-M) (Å) | 13.390 | 13.409 | 13.478 | 13.680 | 13.916 | 13.819 | 13.217 | 13.203 | 13.471 |
| | $J6^{init}$ (Å$^{-1}$) | -0.0412 | -0.0415 | -0.0406 | -0.0393 | -0.0417 | -0.0423 | -0.0425 | -0.0427 | -0.0410 |
| | $j^{init}$ (Å$^{-1}$) | -0.0201 | -0.0201 | -0.0199 | -0.0193 | -0.0203 | -0.0206 | -0.0208 | -0.0209 | -0.0201 |
| | angle MXM | 175.35° | 175.37 | 175.38° | 175.43° | 175.49° | 175.46° | 175.39° | 175.46° | 175.53 |
| | $J6^{displ}$ (Å$^{-1}$) | -0.0413 | -0.0412 | -0.0408 | -0.0395 | -0.0418 | -0.0425 | -0.0434 | -0.0437 | -0.0413 |
| | $j^{displ}$ (Å$^{-1}$) | -0.0202 | -0.0201 | -0.0199 | -0.0193 | -0.0204 | -0.0207 | -0.0208 | -0.0209 | -0.0202 |
| | angle MXM | 175.45° | 175.47° | 175.49° | 175.53° | 175.61° | 175.59° | 175.48 | 175.56 | 175.7° |
| | $J6'^{init}$ (Å$^{-1}$) | -0.0003 | -0.0008 | -0.0005 | 0.0002 | -0.0052 | -0.0053 | -0.0001 | -0.0003 | 0.0003 |
| | $J6'^{displ}$ (Å$^{-1}$) | -0.0003 | -0.0007 | -0.0005 | 0 | -0.0052 | -0.0053 | -0.0001 | -0.0003 | 0.0003 |
| | $y^{init(i)}$ | 0.0010 | 0.0010 | 0.0010 | 0.0010 | 0.0010 | 0.001 | 0.0010 | 0.0009 | 0.0030 |
| | $y^{displ(i)}$ | -0.0010 | -0.0011 | -0.0011 | -0.0011 | -0.0015 | -0.0015 | -0.0010 | -0.0010 | -0.0005 |
| | $d(M\text{-}X)^{init}$ (Å) | 2.393 | 2.389 | 2.396 | 2.416 | 2.499 | 2.491 | 2.389 | 2.391 | 2.436 |
| | $d(M\text{-}X)^{displ}$ (Å) | 2.384 | 2.380 | 2.387 | 2.407 | 2.487 | 2.480 | 2.380 | 2.383 | 2.420 |
| | $d(X\text{-}X)^{init}$ (Å) | 3.19-3.44 | 3.20-3.44 | 3.22-3.46 | 3.28-3.50 | 3.29-3.61 | 3.26-3.58 | 3.12-3.42 | 3.11-3.44 | 3.14-3.53 |
| | $d(X\text{-}X)^{displ}$ (Å) | 3.19-3.44 | 3.20-3.44 | 3.22-3.46 | 3.32-3.51 | 3.29-3.60 | 3.26-3.58 | 3.12-3.42 | 3.10-3.43 | 3.14-3.52 |

(a) The positions of the ions: M – 2c (1/3, 2/3, 1/4); S – 12i (x, y, z); Nb1 – 2a (0, 0, 0); Nb2 – 4f (1/3, 2/3, z); *Z* = 2. Atom positions estimated by editor, not refined

(b) XDP - X-ray diffraction from powder, XDS - X-ray diffraction from single crystal, NDP - neutron diffraction from powder.

[(c)]The refinement converged to the residual factor (*R*) values.
[(d)]Sign and strength of magnetic couplings $Jn^{init}$ and $Jn^{displ}$ ($Jn<0$ – AFM, $Jn>0$ – FM) calculated using the initial and displaced variants of X ions coordinates, respectively.
[(e)]$j^{init}$ and $j^{displ}$ - maximal contributions of the intermediate X ion into the AFM component of the *J*n coupling calculated using the initial and displaced variants of coordinates, respectively.
[(f)]Ratio ($l'/l$) of $l$ and $l'$ segments for the intermediate X ion making the maximal AFM contribution to the $J1_1$ coupling localized near the critical position ‘c’.
[(g)]$\Delta l_y^{displ}$ - displacement of the X ion from the center in parallel to the bond line *Mi –Mj*.
[(h)]$\Delta a=(r_M+r_X)-h_X$ - the distance by which the intermediate X located near the critical position ‘a’ entered the local space between $M_i$ and $M_j$ ions. If $\Delta a<0.1$, the ion does not participate in coupling.
[(i)] $y^{init}$ and $y^{displ}$ - coordinates of the X ion in the initial and mixed variants (*x* and *z* coordinates did not change).

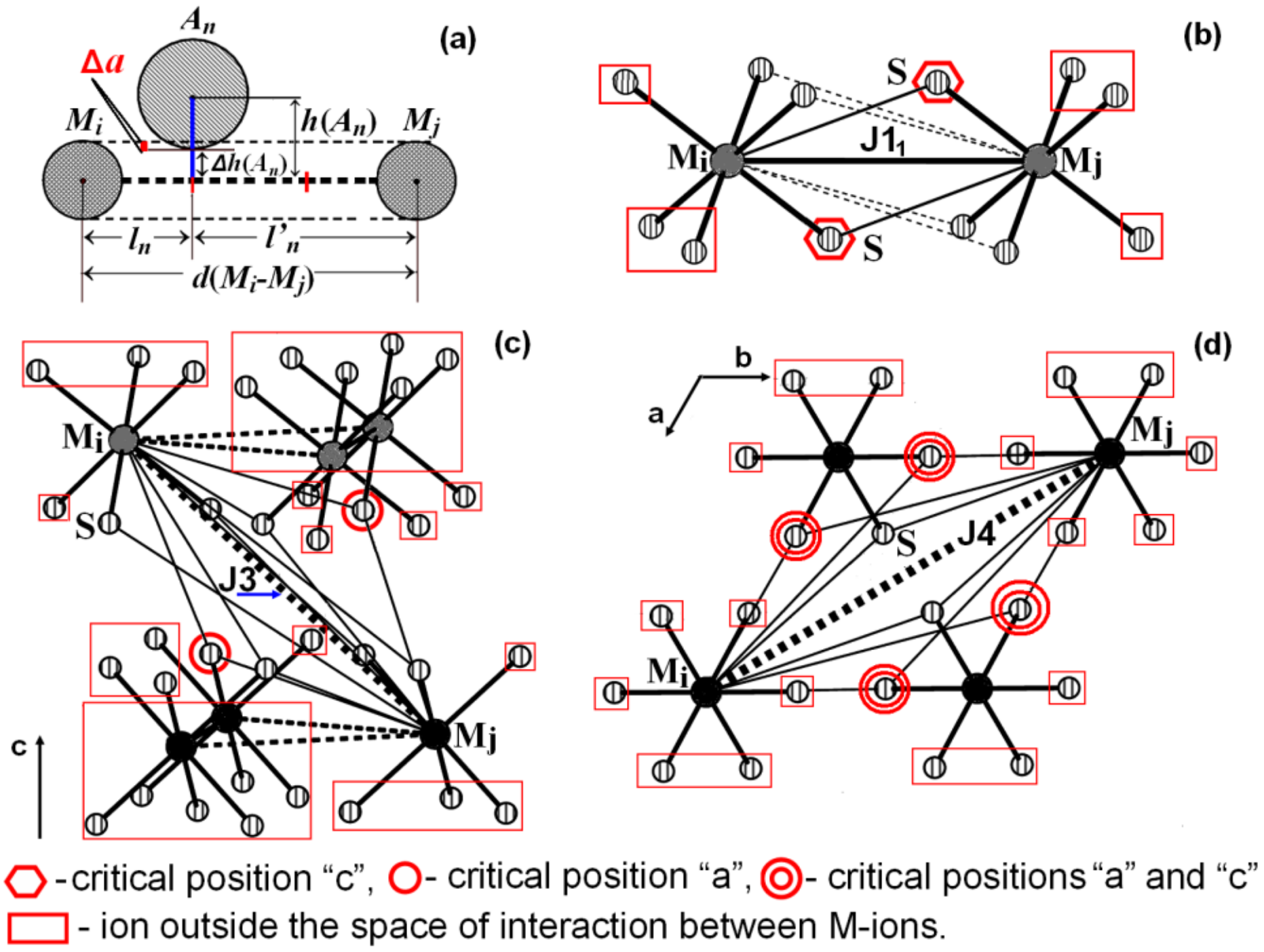

FIG. 3. A schematic representation of the intermediate *An* ion arrangement in the local space between magnetic ions $M_i$ and $M_j$ near two critical positions ‘a’ ($\Delta a \sim 0.1$Å; $\Delta a = (r_M + r_{A_n}) - h_{A_n}$) and 'c' (( $l'_n/l_n = 2$ ) simultaneously; $\Delta h(A_n)$, $l_n$, $l'_n$ and *d(Mi–Mj )* - parameters determining the sign and strength of magnetic interactions (a). The arrangement of intermediate ions in the space of AFM *J*1 (b), FM *J*3 (c) and *J*4 (d) interactions.

strength at insignificant displacement of these S ions to the center along the $M_i$-$M_j$ bond line.

Let us examine changes in the parameters of AFM $J1_1$ couplings exclusively at the expense of displacements of intermediate S ions, which can occur upon the temperature decrease, and check the possibility of transition of these couplings into the FM state within the frames of the space group $P6_322$ with preservation of the structural type $Cr_{1/3}NbS_2$. The latter has certain preconditions: the point is, two intermediate S ions making a substantial contribution to the AFM component of the $J1_1$ coupling are located close to the boundary ($l'/l$ = 1.99, where $l$ and $l'$ are the lengths of segments obtained by drawing a perpendicular from the center of the S ion to the bond line Cr-Cr) of the central one-third of the local space between magnetic ions $Cr^{3+}$ (Fig. 3(a)) and, therefore, to the critical position “c” ($l'_n/l_n$ = 2). In case of an insignificant displacement (by 0.006 Å along the *a* or *b* axis) of S ions from the center in parallel to the bond line Cr-Cr and their emission from the central one-third of the local space, the contribution of each of them to the AFM component of the $J1_1$ coupling would decrease 5-fold (Table I). However, in this case the contribution to the FM component of interaction would not change, and their small values would be of a crucial importance. So the contribution summing up yields that such a displacement would decrease the $J1_1$ coupling value 13-fold. The strength of next-nearest-neighbor $J1_2$ couplings would decrease just 3-fold, while $J1_2$ couplings would undergo the FM→AFM transition and start to compete with weakened $J1_1$ couplings. One should mention that this mixing off S ions would be just insignificantly reflected on the crystal structure. The decrease of Cr-S and S-S distances in the $CrS_6$ octahedron will not exceed 0.01 Å (Table I).

Further reduction of the AFM $J1_1$ coupling strength is possible only at the expense of displacement of the same S ions, making AFM contribution, along the *c* axis. For example, one can achieve the $J1_1$ reduction down to zero, if the S ion (initial coordinates: y = 0.001, z = 0.132, distances in the octahedron d(Cr-S) = 2.393Å) is displaced along the *b* axis by 0.103 Å and along the *z* axis by 0.116 Å (coordinates after displacement: y = -0.017, z = 0.1224, d(Cr-S) = 2.384Å). As a result, the AFM contribution of each of two intermediate S ions would decrease down to -0.004 $Å^{-1}$, whereas summing up these two AFM contributions with four FM ones 0.002 $Å^{-1}$ each would yield their elimination. However, such a displacement would yield the reduction (by 0.228 Å) of the shortest S-S distance between $CrS_6$ octahedra from adjacent layers down to 2.97Å, i.e., to compression of the sandwich layer S-Nb-S.

Nevertheless, it is impossible to implement, in such a way, the transition from AFM $J1_1$ to the FM state, since it is necessary to displace the S ion by too large distance (Δh>0.413 Å). This will yield the destruction of the structural type, because such a displacement would yield compression of the sandwich layer S-Nb-S until distances between S atoms equal to 2.368 Å characteristic for covalently bonded pairs $(S_2)^{2-}$ and elongation of Cr-S bonds until unacceptable values. According to Ref. 38, the minimum value of interionic contact $S^{2-}…S^{2-}$ is equal to 3.07Å.

Thus, the crystal structure of $Cr_{1/3}NbS_2$ induces the emergence of strongly frustrated isotropic AFM triangular planes perpendicular to the *c* axis with competing strong nearest neighbor AFM $J1_1$ couplings. At the same time, it allows, upon temperature decrease, decreasing the strength of $J1_1$ couplings through insignificant displacements of intermediate $S^{2-}$ ions located in critical positions. The decrease of the strength of $J1_1$ couplings would reduce frustration in the triangular plane, which could enable other stronger magnetic couplings available in the stricture to control spin ordering on triangular planes.

Let us consider the effect of magnetic couplings between triangular planes on spin orientations in planes and spin direction at transition from the plane to that along the *c* axis. According to our calculations (Table I), two antiferromagnetic $J2$ ($J2^{init}/J1_1^{init}$ = 0.17, d(Cr-Cr) = 6.897 Å), $J5$ ($J5^{init}/J1_1^{init}$ = 0.32, d(Cr-Cr)= 10.653 Å) and one ferromagnetic $J3$ ($J3/J1_1$ = -0.30, d(Cr-Cr)= 8.974 Å) couplings are available between adjacent triangular planes (Figs. 4(a), 4(b) and 4(c)). The interplane AFM $J2$ and $J5$ couplings compete with the AFM $J1_1$ couplings of triangular planes, since they form AFM triangles $J2J1J2$, $J5J1J5$, and $J2J1J5$. However, this competition is weak, because the $J2$ and $J5$ couplings are weaker than the $J1_1$ one in 6.9 and 3.6 times, respectively.

The picture could change dramatically in case of a substantial decrease of the strength of interplane $J1_1$ couplings at the temperature reduction, as was shown above. Then, under effect of interplane AFM $J2$ and $J5$ couplings that would become stronger ($J2^{displ}/J1_1^{displ}$ = 1.88, $J5^{displ}/J1_1^{displ}$ = 4.15) than $J1_1$ ones, the spin orientation in each plane can be in just one direction (ferromagnetic), while the orientation of adjacent planes - in opposite directions (antiferromagnetic) (Figs. 4(a) and 4(b)). One should mention that such spin orientation is predetermined by the geometry of triangular planes displaced relatively to each other, if magnetic interactions between adjacent planes are antiferromagnetic, in other words, if the structure comprises a system composed of two sublattices (Fig. 2(d)) with opposite spins.

However, in this case, the latter ordering is in competition with the FM $J3$ coupling ($J2^{displ}/J3^{displ}$= -0.46, $J5^{displ}/J3^{displ}$= -1.01), whose impact is directed to rotation of spins of adjacent planes into one direction (Fig. 4(c)). Besides, the transition of $J1_1$ couplings in triangular planes from the AFM into the FM state under effect of interplane couplings $J2$ and $J5$ would induce competition in triangles $J3$(FM)–$J1_1$(FM)–$J2$(AFM). Weak FM $J_c$ ($J_c/J1_1$ = -0.13, d(Cr-Cr) = 12.097 Å = *c*) (Figs. 4(a), 4(b) and 4c)) couplings along the parameter *c* between planes of $Cr^{3+}$ ions through the plane do not compete with interplane $J2$, $J3$, and $J5$ couplings.

It should be emphasized that crystallographic equivalents of interplane couplings $J2$, $J3$, $J5$, and $J_c$ in the $Cr^{3+}$ ions lattice are also magnetic equivalents and do not form chiral helices along the *c* axis. Parameters of these couplings do not change at insignificant displacements of intermediate S ions, unlike AFM $J1_1$ couplings in triangular planes, whose strength could reduce dramatically (see above).

Finally, let us demonstrate which couplings can play a key role in formation of chiral spin structure in $Cr_{1/3}NbS_2$. Such couplings include the strong AFM $J6$ ($J6^{init}/J1_1^{init}$ = 0.77, $J6^{displ}/J1_1^{displ}$ = 10.07, d(Cr-Cr) = 13.390 Å) ones. They emerge between the planes of $Cr^{3+}$ ions through a plane along just one of two crystallographically equivalent diagonals of side faces of the $Cr_7$ prism (Fig. 2(e)) and comprise intertwined left-handed AFM helices twisting along the *c* axis (Figs. 4(d), 4(f) and 4(g)). The contributions to the AFM component of the $J6$ coupling emerge under effect of two S ions (angles Cr-S-Cr = 175.4°), which compose an edge of the chromium octahedron centering the triangular prism $Cr_7$ (Fig. 2(e)).

The AFM magnetic coupling $J6'$ ($J6'^{init}/J1_1^{init}$ = 0.006, $J6'^{displ}/J1_1^{displ}$ = 0.073, $J6'^{init}/J6^{init}$ = $J6'^{displ}/J6^{displ}$ = 0.007, d(Cr-Cr) = 13.390 Å) along another diagonal (Fig. 2(e)) is 140 times weaker than the $J6$ coupling, since S ions do not reach the central one-third of its local space. These couplings could be right-handed, but they are virtually absent. Since in the local space of $J6$ и $J6'$couplings there are no intermediate S ions in critical positions, displacement of these ions does not have any significant effect on either value or sign of these couplings. The helical AFM $J6$ couplings are stronger than the AFM $J2$ ($J2/J6$= 0.22), $J5$ ($J5/J6$= 0.42), FM $J3$ ($J3/J6$= -0.40), and $J_c$ ($J_c/J6$ = -0.17) couplings in 4.6, 2.4, 2.5, and 5.9 times, respectively.

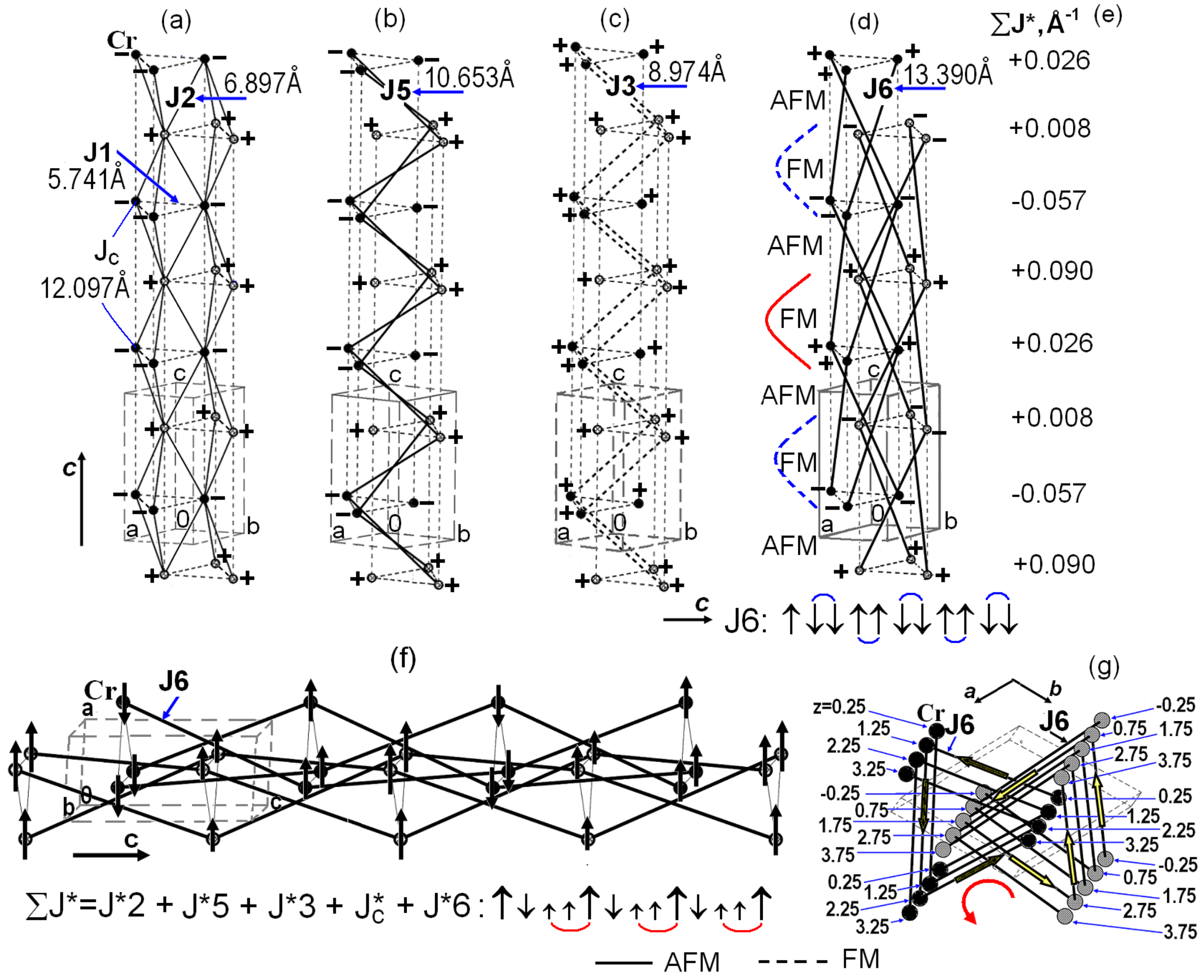

FIG. 4. Spin direction + or - in triangular planes and magnetic ordering of triangular planes along the *c* axis under effect of interplane couplings $J2$ (a), $J5$ (b), $J3$ (c), $J_c$ (a, b, c), and $J6$ (d). The sum ($\sum J^*$) of strengths of interplane couplings $J2$, $J5$, $J3$, $J_c$, and $J6$ orienting magnetic moments of individual triangular planes (e) and the form of magnetic ordering of FM triangular planes along the *c* axis as a result of this summing (f). The view down along [001] of intertwined left-handed helices from $Cr^{3+}$ ions, between which AFM $J6$ couplings emerge (g).

To sum up, we obtain the following picture of ordering of FM triangular planes along the *c* axis under effect of strong helical AFM $J6$ couplings. The $J6$ couplings turn plane spins in sublattices into opposite direction and double the parameter *c* of the magnetic lattice. As a result, magnetic ordering of FM triangular planes along the *c* acquires the form $\uparrow\uparrow\downarrow\downarrow$, i.e., each block composed of two adjacent triangular planes is oriented ferromagnetically, while blocks are ordered in the AFM fashion (Fig. 4(d)). One should mention that the crystal structure allows spin displacement along the helix by a half of period of the elementary unit *c* with ordering principle preservation.

If one formally combines forces of interplane couplings $J2$, $J5$, $J3$, $J_c$, and $J6$ ($\sum J^* = J^*2 + J^*5 + J^*3 + J^*_c + J^*6$) with taking into account their signs (spin direction + or -) in respective planes, the magnetic ordering of FM triangular planes along the *c* axis will have the form $\uparrow\uparrow\uparrow\downarrow\uparrow\uparrow\uparrow\downarrow$ (Figs. 4(e) and 4(f)). The resulting units are blocks composed of three ferromagnetically oriented planes, which are ordered in FM fashion as well and separated by the FM plane with opposite spins. The sum of forces of interplane couplings ($\sum J^*$) orienting magnetic moments of individual planes would change along the *c* axis as follows: +0.008, +0.026, and +0.090 $Å^{-1}$ inside the block and -0.057 $Å^{-1}$ between blocks (Fig. 4(e)). If, in addition, one takes into account the coupling multiplicity ($J2$, $J3$, and $J6$ multiplicities are equal to 6, $J5$ and $J_c$ multiplicities are equal to 12 and 2, respectively), their sum ($\sum J^* = J^*2 + 2J^*5 + J^*3 + J^*_c/3 + J^*6$) will change along the axis in a different waye: +0.020, +0.014, and +0.102 $Å^{-1}$ in block and -0.069 $Å^{-1}$ between blocks.

If crystallographically equivalent $J6$ and $J6'$ couplings were also magnetic equivalents, their contribution into formation of the magnetic structure would be doubled. Then it would turn out that, as a result of summing interplane coupling forces ($\sum J^* = J^*2 + 2J^*5 + J^*3 + J^*_c/3 + 2J^*6$), that the magnetic ordering of triangular planes would almost completely comply with the order determined by these strength-dominating couplings and have the form ↑↑↓↓. Other interplane couplings affect just the value, but not the sign of the sum ∑J**, which changes along the *c* axis as follows: +0.055, +0.144, -0.110, and -0.022 $Å^{-1}$. In spite of formality of this approach, the obtained results indicate to the effect of crystal chirality on its magnetic ordering.

Thus, we demonstrated that at low temperatures the magnetic structure of a layered intercalation compound $Cr_{1/3}NbS_2$ can be transformed into a quasi-one-dimensional system formed by strength-dominant left-handed AFM $J6$ helical chains. These chains are oriented along the *c* axis and packed into two-dimensional triangular lattices in planes perpendicular to the chain direction that lay one above each other at a distance of $c/2$ with a shift at $a/3$ and $b/3$. The magnetic ordering of triangular planes along the *c* axis under effect of strong AFM $J6$ couplings could have the form ↑↑↓↓. However, competition of these AFM helical chains with already weakened, in the course of temperature reduction, frustrated interchain interactions in triangular planes and other weak interplane interactions would facilitate the emergence of a chiral helimagnetic structure in the $Cr_{1/3}NbS_2$ intercalate compound. We assume that in this case the role of the DM interaction consists in final ordering and stabilization of chiral spin helices into chiral magnetic soliton lattice.

Two important factors of the emergence of the left-handed chiral heliomagnetic structure in $Cr_{1/3}NbS_2$ are caused by the crystal structure of this compound: dominating left-handed AFM $J6$ helices and their competition with weaker inter-helix interactions not destructing the system quasi-one-dimensional character. No other chiral helical ways along the *c* axis, where the Cr-Cr distances along the helix are less than 13.4 Å, are available in the magnetic subsystem under examination. A dramatic nonequivalence on the strength of magnetic interactions of left-handed $J6$ and right-handed $J6'$ helices ($J6'/J6$= 0.007), which are crystallographically equivalent, that yields chirality and the possibility of a dramatic reduction of the strength AFM $J1_1$ interactions in triangular planes originate from the noncentrosymmetric location of intermediate $S^{2-}$ ions in the common position 12i.

## IV. POSSIBILITY OF EMERGENCE OF SOLITONS IN INTERCALATION COMPOUNDS $M_{1/3}NbX_2$ AND $M_{1/3}TaX_2$ (M = Cr, V, Ti, Rh, Ni, Co, Fe and Mn; X = S and Se) FROM THE POINT OF CRYSTAL CHEMISTRY

Below we will try to predict, in which representatives of the above family of intercalation compounds, the emergence of solitons is possible. For this purpose, we calculated the parameters of magnetic interactions in 16 more compounds of this type ($M_{1/3}NbS_2$ (M = V, Ti, Ni, Co, Fe and Mn), $M_{1/3}NbSe_2$ (M = Cr, Rh, V, Ti and Co), $M_{1/3}TaX_2$ (M = Cr, V and Rh), and $M_{1/3}TaSe_2$ (M = Cr and V)) and compared them with respective parameters in the soliton $Cr_{1/3}NbS_2$. Table I presents selected room-temperature crystallographic characteristics, parameters of magnetic couplings, and structural data calculated using the initial and displaced variants of coordinates of S(Se) atoms for just 9 of the examined compounds, including $Cr_{1/3}NbS_2$. In the calculations, we used ionic radii of intercalate ions $Cr^{+3}$, $V^{+3}$, $Ti^{+3}$, $Rh^{+3}$ in the trivalent state and $Ni^{+2}$, $Co^{+2}$, $Fe^{+2}$ and $Mn^{+2}$ in the divalent state on two reasons. First, the values of the bond-valence sum (BVS) of intercalate ions (Table I) calculated using bond-valence parameters for anions S and Se[39] are similar to these valence values. Deviations to increasing values can be explained by higher covalent character of metal–chalcogen bonds. Second, these valence states are the most stable for the above intercalate ions in compounds with sulfur and selenium.

In spite of the isostructural character, it turned out that the intercalate compounds under examination were dramatically different with respect to the strength of two magnetic couplings: intraplane $J4$ and interplane $J3$. The point is, four and two intermediate S(Se) ions are located near the boundaries of local spaces of $J4$ (Fig. 3(d)) and $J3$ (Fig. 3(e)) couplings (critical position “a”), respectively. These S(Se) ions do not initiate the emergence of magnetic interaction until their entering inside the local space by at least 0.1 Å ($\Delta a \geq 0.1$Å) (Fig. 3(a)). In the latter case, there emerges a strong FM interaction between magnetic ions. The effect of shifting of intermediate ions inside the local space can be also achieved through the increase of the size of intermediate ion, for instance, through substitution of a small ion (S) by a larger one (Se). Another way consists in the increase of the local interaction space radius at the expense of substitution of small magnetic ions by large ones. That is why in this case one should know the valence state of the magnetic ion, whose radius and, therefore, that of the local space, increase along with the valence decrease. Besides, the FM $J4$ couplings strength could, in addition, be dramatically increased upon shifting of S(Se) ions to the center in parallel to the $M_i$-$M_j$ bond line, since they are simultaneously located near one more critical position "c" ( $l'_n/l_n = 2$ ) (Fig. 3(a)).

The parameters of other couplings ($Jn^{init}$ and $Jn^{displ}$) calculated using the initial and displaced variants of S(Se) ions coordinates, respectively, are variable over strength in rather narrow limits for 17 compounds under examination, except $Mn_{1/3}NbS_2$. The intraplane couplings are within the following limits: AFM $J1_1^{init}$: -0.045 – -0.064 Å$^{-1}$ (-0.025$^{Mn}$ Å$^{-1}$), AFM $J1_1^{displ}$: -0.002 – -0.006 Å$^{-1}$ (0.004$^{Mn}$ Å$^{-1}$, FM), FM $J1_2^{init}$: 0.003 – 0.011 Å$^{-1}$ (0.017$^{Mn}$ Å$^{-1}$), AFM $J1_2^{displ}$: -0.001 – -0.011 Å$^{-1}$ (0.008$^{Mn}$ Å$^{-1}$, FM). Parameters of respective interplane couplings are especially close to each other. Moreover, small displacements of intermediate ions we made to reduce the strength of intraplane couplings $J1_1$ do not virtually change these parameters. The interplane couplings are in the following limits: AFM $J2^{init}$ and AFM $J2^{displ.}$: -0.008 – -0.011 Å$^{-1}$, AFM $J5^{init}$ and AFM $J5^{displ.}$: -0.016 – -0.020 Å$^{-1}$, AFM $J6^{init}$ and AFM $J6^{displ.}$: -0.039 – -0.044 Å$^{-1}$, $J6'^{init}$ and $J6'^{displ.}$: -0.005 Å$^{-1}$ (AFM) – 0.001 Å$^{-1}$ (FM), FM $J_c^{init}$ and FM $J_c^{displ.}$: 0.0005 – 0.008 Å$^{-1}$

The established differences between 17 compounds with respect to the parameters of inteplane ($J3$) and intraplane ($J4$) couplings enable one to divide them into 5 groups. The first group includes four compounds ($Cr_{1/3}NbS_2$[15] (1), $Cr_{1/3}TaS_2$[15] (2), $V_{1/3}TaS_2$[15] (3) and $V_{1/3}NbS_2$[31] (4)) containing small trivalent intercalates ions and sulfur ions. In these compounds, intermediate sulfur ions located in the critical position 'a' remain beyond the boundary of the local space of $J3^{init}$ and $J3^{displ}$ couplings. They enter not too deep ($\Delta a$<0.1 Å) into the local space of $J4^{init}$ and $J4^{displ}$ couplings and do not participate in their formation. As a result, the FM $J3^{init}$ (0.016 Å$^{-1}$), FM $J3^{displ}$ (0.017 Å$^{-1}$), AFM $J4^{init}$ (-0.006 Å$^{-1}$), and и AFM $J4^{displ}$ (-0.005 – -0.006 Å$^{-1}$) couplings in these compounds are significantly weaker than the AFM $J6^{init}$ (-0.041 Å$^{-1}$) and AFM $J6^{displ}$ (-0.041 Å$^{-1}$) couplings in left-handed helices. The parameters of intraplane and interplane couplings in the compounds (2)–(4) of this group are virtually identical to respective couplings of the soliton $Cr_{1/3}NbS_2$ (Table I).

The second group composed of three compounds ($Ti_{1/3}NbS_2$[31] (5), $Cr_{1/3}NbSe_2$[35] (6) и $Cr_{1/3}TaSe_2$[15] (7)) is similar to the first one. Here, as in the first group, S(Se) ions do not enter the local space of the FM $J3^{init}$ (0.011 – 0.017 Å$^{-1}$) and FM $J3^{displ}$ (0.012 – 0.018 Å$^{-1}$) couplings. However, all four S(Se) ions enter the local space of $J4^{init}$ couplings and make FM contributions to their formation. As a result, the $J4^{init}$ (0.001 – 0.007 Å$^{-1}$) couplings are transformed into the FM state. Upon displacement, these 4 S ions leave the local space of the $J4^{displ}$ coupling in the compound $Ti_{1/3}NbS_2$, and the $J4^{displ}$ (-0.007 Å$^{-1}$) coupling becomes antiferromagnetic, just like in the first group of compounds. The displacement of Se ions in $Cr_{1/3}NbSe_2$ and $Cr_{1/3}TaSe_2$ yields a reverse effect – the increase of the strength of FM $J4^{displ}$ couplings (up to 0.016 Å$^{-1}$), in spite of the removal of just two Se ions from the local interaction space. The latter constituted the effect of displacement of the remaining Se ions fromt he second critical position "c" ( $l'_n/l_n < 2$ ).

As follows from the data, the main difference between two groups consists in the fact that the $J4^{init}$ couplings in all three compounds and the $J4^{displ}$ couplings in just two ($Cr_{1/3}NbSe_2$ and $Cr_{1/3}TaSe_2$) of the second group are ferromagnetic, whereas in the first group the couplings are antiferromagnetic and have about the same strength. The differences are hardly capable to exclude the possibility of emergence of solitons in these three compound as well, since the dominating role in the second group belongs to left-handed helices $J6$ upon S(Se) displacements, which dramatically reduce the strength of $J1_1$ couplings in triangular planes.

A different situation takes place in the third, fourth, and fifth groups. Here emerge other leaders surpassing helical AFM $J6$ (-0.040 – -0.043 Å$^{-1}$) couplings in strength. In the third group, including 6 compounds ($Rh_{1/3}NbSe_2$[15] (8), $V_{1/3}NbSe_2$[35] (9), $V_{1/3}TaSe_2$[15] (10), $Ti_{1/3}NbSe_2$[15] (11), $Rh_{1/3}TaS_2$[15] (12), and $Ni_{1/3}NbS_2$[32] (13)), these are the $J4^{displ}$ couplings (0.038 – 0.048 Å$^{-1}$). The reason of a dramatic increase of strengths of these couplings consists in entering (by $\Delta a$>0.1Å) into their local spaces by 4 intermediate S(Se) ions (each) with simultaneous displacement along the M-M bond line until $l'_n/l_n < 2$. Four intermediate S(Se) ions each also entered the local space of the $J4^{init}$ (0.0004 – 0.004 Å$^{-1}$) couplings, but their FM contributions are insignificant, since they were beyond the boundary of the central one-third of the local space ( $l'_n/l_n > 2$ ). Here, just like in groups 1 and 2, the FM $J3^{init}$ (0.011 – 0.016 Å$^{-1}$) and $J3^{displ}$ (0.012 – 0.017 Å$^{-1}$) couplings are relatively weak, since S(Se) ions do not enter by $\Delta a$>0.1 Å into their local space.

In the fourth group containing of two compounds ($Co_{1/3}NbSe_2$[35] (14) and $Co_{1/3}NbS_2$[33] (15)), the strength of the interplane FM $J3$ ($J3^{init}$ and $J3^{displ}$) couplings until was dramatically increased for the first time 0.038$^{Se}$ and 0.047$^{S}$ Å$^{-1}$. Two S(Se) ions each emerged in the local space of the $J3$ coupling due to the increase of its radius upon substitution of small trivalent ions $Cr^{+3}$, $V^{+3}$, $Ti^{+3}$, and $Rh^{+3}$ and divalent ion $Ni^{+2}$ by larger ion $Co^{+2}$. The FM $J3$ couplings are just insignificantly weaker than the AFM $J6$ (-0.043 – -0.044 Å$^{1}$) couplings in $Co_{1/3}NbSe_2$ and stronger than them in $Co_{1/3}NbS_2$. In spite of the fact that 4 S(Se) ions each entered the local space of the $J4$ couplings in both compounds, the FM $J4^{init}$ (0.0001$^{Se}$ and 0.005$^{S}$ Å$^{-1}$) couplings are significantly weaker than the FM $J4^{displ}$ (0.019$^{Se}$ and 0.027$^{S}$ Å$^{-1}$), since in the first case the FM contribution to interaction of all four intermediate ions is small ( $l'_n/l_n > 2$ ), while in the second case the contribution of two of them is large ( $l'_n/l_n < 2$ ).

In the compounds of the fifth group ($Fe_{1/3}NbS_2$[34] (16) and $Mn_{1/3}NbS_2$[32] (17)) with large magnetic ions $Fe^{+2}$ and $Mn^{+2}$, the interplane FM $J3^{init}$ ($0.047^{Fe}$ Å$^{-1}$ and $0.045^{Mn}$ Å$^{-1}$) and $J3^{displ}$ ($0.048^{Fe}$ and $0.046^{Mn}$ Å$^{-1}$) and intraplane intraplane FM $J4^{displ}$ ($0.055^{Fe}$ and $0.067^{Mn}$ Å$^{-1}$) couplings are stronger than the AFM $J6$ ($-0.041^{Fe}$ and $-0.040^{Mn}$ Å$^{-1}$) couplings in left-handed helices. The reasons of strengthening of the FM $J3$ and FM $J4^{displ}$ couplings are teh same like in the fourth and thrid groups, respectively.

We believe that the dramatic change in the strength of just one coupling ($J3$ or $J4$) excludes, along with the dominating of AFM $J6$ helical couplings, the possibility of soliton emergence as well. That is why only in the compounds of the first two groups ($Cr_{1/3}TaS_2$, $V_{1/3}TaS_2$, $V_{1/3}NbS_2$, $Ti_{1/3}NbS_2$, $Cr_{1/3}NbSe_2$ and $Cr_{1/3}TaSe_2$), in which the dominating role belongs to the AFM $J6$ couplings of left-handed helices, whereas the couplings between helices are weak, we expect the emergence of solitons similar to those existing in $Cr_{1/3}NbS_2$.

## V. CONCLUSIONS

The role of structural factors in the formation of chiral magnetic soliton lattice in the intercalation compound $Cr_{1/3}NbS_2$ has been considered. Analysis of the parameters of magnetic interactions calculated using the crystal chemistry method on the basis of structural data has demonstrated that two important factors of formation of chiral magnetic soliton lattice in $Cr_{1/3}NbS_2$ are caused by the compound crystal structure. These factors include the dominating left-handed AFM $J6$ helices and their competition with weaker inter-helix AFM interactions resulting in a non-collinear helical magnetic structure not destructing the quasi-one-dimensional character of the magnetic subsystem.

The latter is expressed, first, in a dramatic nonequivalence of the strengths of crystallographically equivalent AFM $J6$ and $J6'$ magnetic interactions ($J6'/J6 = 0.007$) between triangular planes of $Cr^{3+}$ ions through that plane, which form left-handed $J6$ and right-handed $J6'$ helices along the $c$ axis. The reason if chirality here is the noncentrosymmetrical position of intermediate $S^{2-}$ ions. The loss of the inversion center in the magnetic subsystem allows manifestation of relativistic forces described by Dzyaloshinskii–Moriya.

Second, this is expressed in a dramatic reduction of the strength of the AFM $J1_1$ couplings in triangular planes at the expense of an insignificant displacement of intermediate $S^{2-}$ ions located in the critical position in the local space of these interactions. As a result, the frustration weakens and the magnetic system becomes quasi-one-dimensional.

Comparison of the parameters of magnetic interactions calculated on the basis of structural data in intercalation compounds $M_{1/3}NbX_2$ and $M_{1/3}TaX_2$ (M = Cr, V, Ti, Rh, Ni, Co, Fe and Mn; X = S and Se) with respective parameters in the soliton $Cr_{1/3}NbS_2$ shows that the emergence of solitons is possible only in the compounds $Cr_{1/3}TaS_2$, $V_{1/3}TaS_2$, $V_{1/3}NbS_2$, $Ti_{1/3}NbS_2$, $Cr_{1/3}NbSe_2$, and $Cr_{1/3}TaSe_2$.

We believe that, through using the Inorganic Crystal Structure Database (ICSD) and the crystal chemistry method, it is possible to find among noncentrosymmetric magnetic compounds candidates for the role of solitons having two mentioned attributes: dominating chiral helices and competition with weaker inter-helix interactions not destructing the system quasi-one-dimensional character. Here, the crystal structures of potential solitons can be of different types.

## ACKNOWLEDGMENTS

This work was financially by the grant 12-I-P8-05 of the programs of fundamental research of the Presidium of the Russian Academy of Sciences.